\documentclass[twocolumn,floatfix]{aastex62}

\usepackage{savesym}
\savesymbol{tablenum}
\usepackage{siunitx}
\usepackage{bm}
\usepackage{mathtools}

\usepackage{color}
\definecolor{midblue}{RGB}{44,133,255}
\definecolor{navy}{RGB}{35,46,127}
\definecolor{tmp}{rgb}{0,.7,0.5}

\graphicspath{{./}{figures/}}

\received{\today}
\revised{XXXX}
\accepted{XXXX}
\submitjournal{ApJ}

\shorttitle{Reionization and Binary Evolution}
\shortauthors{Secunda et al.}

\newcommand{\citeg}[1]{\citep[e.g.,][]{#1}}
\newcommand{\citesee}[1]{\citep[see][]{#1}}

\begin{document}

\title{Delayed Photons from Binary Evolution Help Reionize the Universe}

\correspondingauthor{Amy Secunda}
\email{asecunda@princeton.edu}

\author{Amy Secunda}
\affil{Department of Astrophysical Sciences, Princeton University, Peyton Hall, Princeton, NJ 08544, USA}

\author{Renyue Cen}
\affiliation{Department of Astrophysical Sciences, Princeton University, Peyton Hall, Princeton, NJ 08544, USA}

\author{Taysun Kimm}
\affiliation{Department of Astronomy, Yonsei University, 50 Yonsei-ro, Seodaemun-gu, Seoul 03722, Republic of Korea}

\author{Ylva G\"{o}tberg}
\affiliation{The Observatories of the Carnegie Institution for Science, 813 Santa Barbara Street, CA 91101 Pasadena, USA}

\author{Selma E.\ de Mink}
\affiliation{Center for Astrophysics, Harvard \& Smithsonian, 60 Garden Street, Cambridge, MA 02138, USA}

\begin{abstract}
High-resolution numerical simulations including feedback and aimed at calculating the escape fraction ($f_{\rm{esc}}$) of hydrogen-ionizing photons often assume stellar radiation based on single-stellar population synthesis models. However, strong evidence suggests the binary fraction of massive stars is $\gtrsim 70\%$. Moreover, simulations so far yield values of $f_{\rm{esc}}$ falling only on the lower end of the $\sim 10-20\%$ range, the amount presumed necessary to reionize the Universe. Analyzing a high-resolution (4 pc) cosmological radiation hydrodynamic simulation we study how $f_{\rm{esc}}$ changes when we include two different products of binary stellar evolution -- stars stripped of their hydrogen envelopes and massive blue stragglers. Both produce significant amounts of ionizing photons 10-200~Myr after each starburst. We find the relative importance of these photons are amplified with respect to escaped ionizing photons, because peaks in star formation rates (SFRs) and $f_{\rm{esc}}$ are often out of phase by this 10-200~Myr. Additionally, low mass, bursty galaxies emit Lyman continuum radiation primarily from binary products when SFRs are low. Observations of these galaxies by the James Webb Space Telescope could provide crucial information on the evolution of binary stars as a function of redshift. Overall, including stripped stars and massive blue stragglers increases our photon-weighted mean escape fraction ($\langle f_{\rm{esc}}\rangle$) by $\sim$13\% and $\sim$10\%, respectively, resulting in $\langle f_{\rm{esc}}\rangle=17\%$. Our results emphasize that using updated stellar population synthesis models with binary stellar evolution provides a more sound physical basis for stellar reionization.

\end{abstract}


\section{Introduction}
\label{sec:intro}

High-redshift, star-forming, dwarf galaxies with virial masses that range from $10^8-10^{10.5}$~M$_{\rm{\odot}}$ are the most plausible source of the hydrogen-ionizing radiation responsible for the reionization of the Universe by z $\sim$ 6, provided a high enough escape fraction ($f_{\rm{esc}}$) \citep{2001_haehnelt, 2009_cowie_barger_trouille,2014_fontanot,2017_madau_fragos,2015_madau_haardt}. Models of cosmic reionization \citeg{2012_kuhlen_FG,2015_robertson} show that $f_{\rm{esc}} \gtrsim 20\%$ is required for cosmic reionization. The Thompson opticial depth \citet{2012_kuhlen_FG} calibrated their results against was a bit higher than current estimates by Planck, and the exact requirement for $f_{\rm{esc}}$ depends on uncertain parameters such as the clumping factor of the intergalactic medium (IGM) and the intrinsic ionizing luminosity density of the early Universe. Nonetheless, $f_{\rm{esc}}$ 
$\ge 10-20\%$ is probably required.

However, observations of Lyman continuum (LyC) in the local Universe, which is limited to starburst galaxies, generally suggest low escape fractions of $\lesssim$ 8\% \citep{2011_leitet,2013_leitet,2014_borthakur,2016_leitherer,2016_izotov}. Star-forming galaxies at $z\sim 1$ with LyC detections also show low escape fractions of a few percent \citep{2007_siana,2010_siana,2010_bridge,2016_rutkowski}. Although observations at redshifts z $\gtrsim 3$ \citeg{2016_reddy,2016_leethochawalit,2018_steidel} seem to suggest that galaxies with $f_{\rm{esc}} \gtrsim 10\%$ are more common than at lower redshifts, only a few of these galaxies with high values of $f_{\rm{esc}}$ have been shown to be robust detections uncontaminated by foreground sources \citep{2015_mostardi,2016_shapley}. 
Interestingly, the average relative escape fraction, the ratio of the escape fraction of $990~\rm{\AA}$ to $1500~\rm{\AA}$ photons, is found to be small (most often $\lesssim 5\%$) in nearly all observations, even at high redshift \citep{2010_vanzella, 2011_boutsia, 2015_mostardi, 2015_siana, 2015_grazian,2017_marchi}, see however, \cite{2019_rivera-thorsen}.

Theory also struggles to provide a value of $f_{\rm{esc}}$ $\gtrsim 10-20\%$. Recent high resolution ($\lesssim$ 10pc) numerical simulations that include feedback \citeg{2014_kimm,2015_ma,2016_xu} suggest average escape fractions of only $5-11\%$. In these simulations the escape fraction is lowest during periods of high star formation (i.e. the periods in which LyC is most often observed in the local Universe) before feedback has the opportunity to clear the LyC-trapping gas from the birth clouds of the stars. 

\cite{2014_wise} suggested that ionizing photons from low luminosity (M$_{\rm{UV}}>-13$) mini-halos could play a large role in reionization. However, \cite{2017_kimm} found that although the escape fractions of these mini-halos are generally large ($\sim20-40\%$), the inefficiency of star-formation in these mini-halos means they play only a minor role in reionization. Using a simple analytic argument, \cite{2012_conroy} showed that including runaway O/B stars may enhance $f_{\rm{esc}}$ by a factor of up to $\sim4.5$ for halos with virial masses between $10^8$ and $10^9$~M$_{\rm{\odot}}$. However, \citet{2014_kimm} and \citet{2015_ma} found that including a simple model for runaway O/B stars in their simulations, only increased their mean values of $f_{\rm{esc}}$ to $14\%$ and $\sim6\%$, respectively.

One of the most important theoretical findings on dwarf galaxies at high redshift is that star formation is very episodic.
Peaks in star formation and $f_{\rm{esc}}$
tend to be out of phase by $10-30$~Myr,
with the latter lagging the former \citep{2014_kimm}.
When star formation is most vigorous in dense regions, the produced ionizing photons do not easily escape.
On the other hand, when supernova feedback has cleared out channels in the ISM,
the O/B stars produced at the peak of star formation
that dominate LyC production \citep[e.g.,][]{1999_leitherer} are gone.

Thus, potentially the most promising additional source of H-ionizing radiation are stars that interact in binaries. Spectroscopic observations of young massive stars in the Milky Way and Magellanic Clouds, show that a very large fraction ($\gtrsim 0.7$) of stars are part of close binary systems \citeg{2007_kobulnicky_fryer, 2009_mason, 2012_sana, 2012_chini,2017_moe,2017_almeida}. In these binary systems interactions between the stars can lead to the exchange of mass and angular momentum through Roche-lobe overflow, common envelope evolution, and the merging of the two stars \citep{1967_kippenhahn_weigert,1976_paczynski,1999_wellstein_langer,2011_ivanova,2013_de_Mink, 2016_schneider, 2017_de_marc_izzard}. These interactions can increase the number of high-mass stars at later times and can create envelope-stripped helium stars, which both emit ionizing photons tens to hundreds of Myr after a starburst \citep{1999_vanbever,2019_gotbergA}. These ``delayed" ionizing photons could be particularly effective in increasing $f_{\rm{esc}}$ since it would give time for feedback from massive stars to remove most of the surrounding gas from the birth cloud that would normally trap LyC radiation \citep{2019_gotbergB}.

\citet[hereafter M16]{2016_ma} studied the effect binary stellar evolution had on the escape fractions of three example halos from the Feedback in Realistic Environment project \citep[{\sc fire},][]{2014_hopkins} using models from the Binary Population and Spectral Synthesis code \citep[{\sc bpass},][]{2008_eldridge,Eldridge_2009,2011_eldridge,Stanway_2016,2017_eldridge}, which accounts for the mass transfer and mergers of a binary star population. They found that while their stellar population synthesis model that did not include binary stellar evolution produced an $f_{\rm{esc}}$ below 5\% at most redshifts for their example halos, the inclusion of binaries in their stellar population synthesis model increased $f_{\rm{esc}}$ by factors of $\sim 3-5$. Including binaries in their model also increased the amount of ionizing photons by a factor of 1.5, leading to a factor of $\sim 4-10$ increase in the ``effective" escape fraction. \citet[hereafter R18]{2018_rosdahl}, also used a model from {\sc bpass} that includes binary stellar evolution in their {\sc sphinx} suite of cosmological adaptive mesh refinement simulations. R18 found a similar increase in their photon-weighted mean escape fraction by a factor of $\sim 3$ from $f_{\rm{esc}} \sim 2-3\%$ for a single-stellar population synthesis model, to $f_{\rm{esc}} \sim 7-10\%$ for their run that included binary stellar evolution.

In this paper we focus on, separately, two different products of interactions between stars in binary systems, namely, envelope-stripped helium stars (see 
\S \ref{sec:method:binary:helium}) and stars that gain mass via mass exchange in binaries and become massive blue stragglers (see \S \ref{sec:method:binary:mergers}).
A significant advantage of this approach is that we can investigate the effect on the escape fraction from different types of binary products, which allows us to better understand which sources are responsible for any change in the escape fraction. Equally important, it is easier to see how the uncertainties in the different aspects of our models produce uncertainties in the emission rate of ionizing photons.

We perform post-processing on the cosmological radiation hydrodynamic simulations from \citet{2014_kimm} (hereafter, KC14) using the ionizing photon rates of stripped stars from \citet{2019_gotbergA}, and a simple model for massive blue stragglers. Our aim is to better understand the role of stripped stars and massive blue stragglers during reionization and their affect on the escape fraction. 
 The outline of the paper is as follows:
 in \S \ref{sec:method:cosmo} we briefly describe the cosmological simulations used. In 
 \S \ref{sec:method:binary} we describe the implementation of the various stellar populations included in our model, including massive stars (\S \ref{sec:massive_stars}), stripped stars 
 (\S \ref{sec:method:binary:helium}), blue stragglers 
 (\S \ref{sec:method:binary:mergers}), and runaway stars 
 (\S \ref{sec:method:runaways}). In \S \ref{sec:methods:binary:other_sources} we discuss other sources of ionizing radiation that we have not included in our simulations, and in \S \ref{sec:method:binary:bpass} we compare our stellar population synthesis models to the models M16 and R18 used from {\sc bpass}. In \S \ref{sec:method:ray} we describe our calculation of $f_{\rm{esc}}$. In 
 \S \ref{sec:results} we show our results for a larger mass example halo (\S \ref{sec:results:one_halo}) and all of the halos in our simulation (\S \ref{sec:results:all}). 
 In \S \ref{sec:compare} we compare our results with two previous studies, and in \S \ref{sec:discuss} we summarize our results.

\section{Method}
\label{sec:method}

\subsection{Cosmological Simulations}
\label{sec:method:cosmo}
We perform our calculation of the escape fraction through post-processing of the ``FRU" cosmological simulations of
KC14, generated using {\sc ramses}
cosmological AMR code \citep{2002_teyssier}.
This enables us to make direct comparisons 
to the results presented in KC14 that
do not include the two effects due to binary stellar evolution. Here we briefly summarize the key components of these simulations.

KC14 use the {\sc music} software \citep{2011_hahn_abel} to generate the initial conditions with the WMAP7 cosmological parameters \citep{2011_komatsu}: ($\Omega_{\rm{m}}$, $\Omega_{\rm{\Lambda}}$, $\Omega_{\rm{b}}$, $h$, $\sigma_{\rm{8}}$, $n_{\rm{s}}$ =0.272, 0.728, 0.045, 0.702, 0.82, 0.96).
They 
first run a dark matter only simulation with 
a sufficiently large box of size (25 Mpc$h^{-1}$)$^3$ with 256$^3$ dark matter particles in order to 
sample the large-scale gravitational field 
and include effects from the large-scale tidal field
in the zoom-in simulation in the next step.

Next, a zoom-in region of comoving size of $3.8 \times 4.8 \times 9.6$~Mpc (comoving) 
is chosen,
where a finer spacing is implemented 
in the initial condition to achieve 
a dark matter particle mass resolution of \num{1.6e5}~$\rm{M_{\odot}}$, effectively corresponding to 2048$^3$ 
over the entire box. 
The zoomed-in region is
dynamically refined according to a pre-set 
criterion to better resolve the structure of the ISM, with up to 12 additional levels of refinement. This refinement
results in a minimum physical cell size of 4.2~pc, and a stellar mass resolution of approximately 49~$\rm{M_{\odot}}$. The {\sc amiga} halo finder \citep{2004_gill,2009_knollmann_knebe} was used to identify dark matter (sub) halos and galaxies.

The adaptive mesh refinement (AMR) code, {\sc ramses} \citep{2002_teyssier} is based on the fully threaded oct-tree structure \citep{1998_khokhlov}, and uses the second-order Godunov scheme to solve the Euler equations. The hydrodynamic states reconstructed at the cell interface are limited using the MinMod method, and then advanced using the Harten-Lax-van-Leer contact wave Riemann solver \citep{1994_toro}. KC14 use a typical Courant number of 0.8 and solve the Poisson equations using the adaptive particle-mesh method.
Star formation and stellar feedback from supernova (SN) explosions are included in these simulations as outlined in KC14. Also included are radiative cooling \citep{1993_sutherland_dopita,1995_rosen_bregman} and thermal stellar winds as in {\sc starburst99} \citep{1999_leitherer}.

KC14 also use the multi-group radiative transfer (RT) module developed by \cite{2013_rosdahl} to 
follow radiative transfer of ionizing photons from  stars through the ISM and the IGM. The module solves the moment equations for HI, HeI, and HeII ionizing photons using a first-order Godunov method with M1 closure for the Eddington tensor. KC14 adopt a Harten-Lax-van-Leer \citep{1983_harten} intercell flux function, and use a photon production rate corresponding to a Kroupa IMF \citep{2001_kroupa} from the {\sc starburst99} library \citep{1999_leitherer}.

\subsection{The ionizing emission from a stellar population}
\label{sec:method:binary}

\begin{figure}
    \centering
    \includegraphics[width=0.45\textwidth]{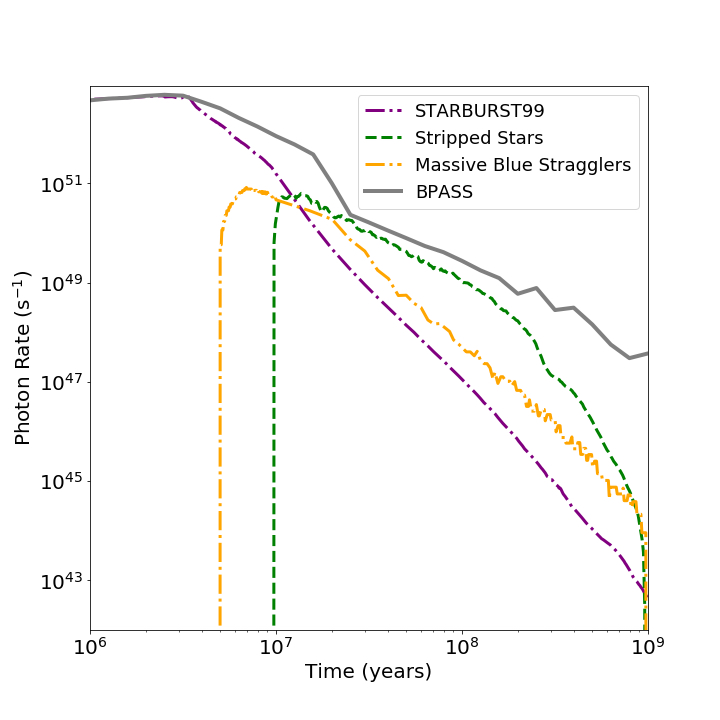}
    \caption{The photon production rate [s$^{-1}$] of LyC photons for a single $10^6$~M$_{\rm{\odot}}$ starburst for the massive star component (in purple), stripped star component (in green), and massive blue straggler component (in orange). For comparison, we also include the LyC photon production rate of the model including binary stellar evolution that M16 and R18 used from {\sc BPASS} \citep[version 2.0]{2017_eldridge,Stanway_2016}. We show the photon production rates for the lowest metallicity available for each population. In our simulation a star particle with a metallicity $\lesssim \num{2e-4}$ would be given the production rate shown here. After about 10~Myr the photon production rate of the stripped stars becomes the dominant photon rate. The photon production rate of the massive blue stragglers peaks at about 5~Myr after the initial starburst and becomes larger than the photon production rate of the {\sc starburst99} component at around 15~Myr. Including photons from stripped stars increases the total amount of ionizing emission by 3.6\% and including photons from massive blue stragglers increases the total amount of ionizing emission by 3.3\%.}
    \label{fig:single_sb}
\end{figure}

In addition to massive stars from a single-stellar population syntheses model, we account for two types of products of binary interaction: stars stripped of their hydrogen-rich envelopes (see 
\S \ref{sec:method:binary:helium}) and stars that gain mass both from mass transfer and from coalescence (see \S \ref{sec:method:binary:mergers}). The stripped stars are the exposed hot and compact cores of their progenitors. In cases where their progenitors were high-mass stars, the stripped stars are Wolf-Rayet stars. The mass-gainers are rejuvenated and therefore appear bluer than the rest of the population. They are therefore often referred to as massive blue stragglers \citeg{1999_vanbever,2009_chen_han,2010_chen_han}.
 
Note that we are replacing the stellar population synthesis models in post-processing, whereas the photoionization in the cosmological simulations is calculated using only the {\sc starburst99} models (see \S \ref{sec:method:cosmo}). If the photoionization in the cosmological simulation was calculated based on a stellar population that included binary stars, photons from these binary stars that failed to escape their host galaxy would ionize the gas that captures them. If these photons were able to ionize gas it would increase the escape fraction in our simulations. We plan to include these effects on the fly in our future simulations.

The emission rates of hydrogen-ionizing photons from massive stars, stripped stars, and massive blue stragglers for a starburst of initially $10^6$~M$_{\odot}$ are shown as a function of time in Fig.~\ref{fig:single_sb}. We also show the ionizing photon emission rates from the model from {\sc bpass} that includes binary stellar evolution that M16 and R18 used, for comparison \citep[version 2.0]{Stanway_2016,2017_eldridge}. Because most star particles in our simulation of high-redshift galaxies are very metal poor, we show the ionizing photon emission rates for the lowest metallicity available for each stellar population. The lowest metallicity of each population is $Z=$ \num{2e-4} for the stripped stars, $Z=$ \num{4e-4} for the massive single stars and massive blue stragglers, and $Z=$ $10^{-3}$ for the model from {\sc bpass}. In our simulation a star particle with $Z \lesssim \num{2e-4}$ would be given the emission rates shown in Fig.~\ref{fig:single_sb}.

Each star particle in the {\sc ramses} simulation has an age and metallicity. For each stellar population we interpolate between the different available ages and metallicities to assign an ionizing photon production rate to each star particle. Below, we describe each model used in our simulations for the individual types of stars in more detail.

\subsubsection{Massive stars}
\label{sec:massive_stars}

We model the ionizing emission from massive stars using {\sc starburst99} \citep{1999_leitherer,2014_leitherer}. We use the stellar models from Padova \citep{1993_bertelli,1994_bertelli,marigo_2008}, {\sc cmfgen} \citep{1998_hillier_miller} and WM-basic \citep{pauldrach_2001} atmospheric models, and assume a Kroupa \citep{2001_kroupa} initial mass function that stretches from 0.1 up to 100 M$_{\odot}$. We use the metallicities $Z = 0.02$ (solar), 0.008, 0.004, and \num{4e-4} that are available in {\sc starburst99}. 

The dominating source of ionizing photons in {\sc starburst99} are the massive O/B-type main sequence stars. The most massive stars die after a few Myr. Their deaths lead to the decline in the emission rate of ionizing photons seen in Fig.~\ref{fig:single_sb}, as the remaining stars are cooler and less luminous making them less efficient ionizing sources. 

Although {\sc starburst99} is created for stellar populations containing only single stars, it does a reasonable job representing the emission from a population that also contains binary stars on the main sequence. It does a reasonable job because binary interaction primarily occurs when the most massive star in the system has depleted hydrogen in its center and is expanding to become a red (super)giant \citep{de_Mink_2008}. During this phase the star's radius expands significantly more than during the star's previous main sequence evolution making interactions with companions more likely to occur. However, mass gain through accretion or coalescence can affect main-sequence stars as well. For example, we do not subtract the contribution to the binary product from the remaining main-sequence star, meaning that the contribution from {\sc starburst99} is somewhat overestimated. However, the binary product from these processes is a significantly brighter and hotter main-sequence star (see \S \ref{sec:method:binary:mergers}), and we therefore consider the overestimate of ionizing emission from {\sc starburst99} to be negligible.

Our default calculation uses only the ionizing emission predicted by {\sc starburst99}. We refer to this simulation as the SB99 run.

\subsubsection{Stars stripped in binaries}
\label{sec:method:binary:helium}

We account for the ionizing emission from stripped stars in stellar populations of different metallicities by adopting the emission rates of ionizing photons presented by \citet{2019_gotbergA} \footnote{ see also the online interface for {\sc starburst99} \url{http://www.stsci.edu/science/starburst99/docs/default.html}}.  \citet{2019_gotbergA} calculated the number and type of stripped stars present in stellar populations as a function of time. Convolving with the emission rate of ionizing photons from individual stripped stars of given masses \citep{2018_gotberg}, they then computed the total emission rate of ionizing photons from stripped stars in stellar populations. The models are created using an initial mass function from \citet{2001_kroupa} with lower and upper mass limits of 0.1 and 100~M$_{\odot}$ and for the metallicities $Z = 0.014$, 0.006, 0.002, and \num{2e-4}. 

In Fig.~\ref{fig:single_sb}, the stripped stars (green, dashed curve) start contributing LyC emission about 10~Myr after a starburst, and become the dominant source of LyC photons soon after. This delay corresponds to the main sequence lifetime of a 20~M$_{\odot}$ star, which is the most massive and thus shortest lived star that \citet{2019_gotbergA} consider as progenitors for stripped stars. Over 1~Gyr stripped stars at this metallicity ($Z$ = \num{2e-4}) contribute an additional 3.6\% of the total LyC emission from massive stars. Stripped stars contribute a higher percentage of the total number of LyC photons at higher metallicities. For example, at solar metallicity over 1~Gyr stripped stars contribute an additional $\sim$6\% of the total LyC emission from massive stars. The relative size of the stripped star contribution changes because, while massive stars emit fewer ionizing photons at higher metallicities, in our models the emission rates of the stripped star stellar population remains reasonably constant as a function of metallicity.

In one of our simulations, we account for the ionizing emission from massive stars and stars stripped in binaries. We refer to this run as the SSB run. 

\subsubsection{Massive blue stragglers}
\label{sec:method:binary:mergers}

We take a simple approach when estimating the ionizing contribution from stars that accrete mass or merge during interaction with a binary companion. 
During mass transfer, the accreting star evolves on its main sequence. When it gains mass from the donor star, it gets rejuvenated as the convective core grows into the hydrogen-rich layers when the star increases in mass. This means that the emission rate of ionizing photons from a star that has accreted mass can be modeled as a younger and more massive main sequence star. 

The result of a merging binary star is most probably a more massive and rejuvenated main sequence star in the case in which the merger occurred during the main sequence evolution of both stars \citep{1997_tout,2019_schneider}. If the merger occurred when one of the stars is in a later evolutionary phase, the outcome of the merger is uncertain, but it is possible that the star expands and becomes a red supergiant. Since red supergiants are inefficient emitters of ionizing photons, we only account for mergers that occur during the main sequence evolution of the two stars in the binary. We take the same approach when modeling the ionizing emission from blue stragglers resulting from mass accretion and coalescence. 

First, we assume that the interaction occurs when the main sequence lifetime of the most massive star in the system has passed. Second, we assume that the blue straggler is rejuvenated such that it is halfway through its main sequence evolution after the rejuvenation. Third, we assume that the blue straggler becomes 50\% more massive than what the most massive star in the system was initially. Last, we assume that 10\% of the stars with initial primary masses between 2 and 50 M$_{\rm{\odot}}$ in a population go through this evolutionary phase and become blue stragglers. 

The ionizing emission in {\sc starburst99} comes primarily from main sequence stars, but also with some contribution from Wolf-Rayet stars. It is a reasonable assumption that mass gaining stars evolve in similar ways as single stars, at least for the main sequence duration. We therefore use the predictions from {\sc starburst99} to estimate the emission rates of ionizing photons from mass gaining stars. It is done as follows. Using lifetimes of massive stars from evolutionary models \citep[see e.g., Fig. 1 of][]{2017_zapartas1}, we infer what the emission rate of ionizing photons from each mass bin of stars is via interpolation and subtracting the contribution from lower mass stars. We calculate what the contribution from each mass bin should be following the assumptions described above. We then shift it with the time delay corresponding to the time of interaction. 

The resulting emission rate of LyC photons can be seen in Fig.~\ref{fig:single_sb}, marked with a dash-dotted, yellow line. The mass-gaining stars begin contributing ionizing photons after several Myr have passed. After around 20~Myr the ionizing contribution of these mass gainers becomes more significant than that of the stars in the population that did not interact. Since the mass gainers are more massive than the single stars that are present in the stellar population, they also mildly harden the ionizing emission from the stellar population. Over 1~Gyr at a metallicity of $Z$ = \num{4e-4} our blue straggler population contributes a total of 3.3\% of the ionizing radiation contributed by the single-stellar population. The contribution from blue stragglers decreases slightly at higher metallicities. For example, at solar metallicity blue stragglers contribute a total of 2.1\% of the ionizing radiation contributed by the single-stellar population.

We refer to the run where we include ionizing radiation from massive stars and massive blue stragglers as MBS.

\subsubsection{Runaway Stars}
\label{sec:method:runaways}
\cite{2011_tetzlaff} found that roughly 30\% of O/B stars are runaway stars with peculiar velocities greater than 28 km~s$^{-1}$. They found that these peculiar velocities fall along a Maxwellian distribution with a dispersion of 24.4~km~s$^{-1}$. Motivated by \citet{2011_tetzlaff}, in their ``FRU" simulations KC14 divided 30\% of their star particles into runaway O/B stars. They drew the peculiar velocities of these runaway stars from a Maxwellian distribution with a dispersion of 24.4~km~s$^{-1}$ and a minimum space velocity of 28 km~s$^{-1}$ and added them to the initial velocity of the star particle. The directions of motion of runaway stars were chosen at random. 

The star particles which represent runaway stars are given the same ionizing photon rates as the other star particles, including the photon rates for  stripped stars and massive blue stragglers in the SSB and MBS  runs. There are currently two proposed channels for the formation of runaway stars. The first way a star could become a runaway is through a three-body interaction with other stars in a young cluster \citep{1988_leonard}, and the second way is from a SN explosion of a companion in a binary system \citep{1961_blaauw}. The second channel would likely result primarily in runaway massive blue stragglers or main-sequence stars, depending on the level of interaction between the stars in the binary system before ejection. It could also potentially result in a runaway stripped star \citep{1994_pols,Renzo_2019}. \citet{Renzo_2019} even found that stripped stars are just as likely to become runaways as massive blue stragglers. On the other hand, if fewer stripped stars become runaways, then fewer stripped stars will end up towards the outskirts of a galaxy where their emission would more easily be able to escape into the IGM. Therefore, fewer runaway stripped stars would result in a lower escape fraction. However, because the first channel for the creation of runaway stars would be impartial to stellar type, and because the relative importance of each channel is still not well constrained \citeg{2000_hoogerwerf, 2001_hoogerwerf, 2005_guseinov}, we consider this simple approach adequate.

\subsubsection{Other Sources of Ionizing Radiation}
\label{sec:methods:binary:other_sources}
There are other stellar sources of ionizing radiation that we do not yet account for, such as post-AGB stars \citep{2015_stasinska,2019_byler}, accreting white dwarfs \citep{1992_van_den_heuvel,2013_woods,2014_woods,2015_chen}, and X-ray binaries \citep{2013_fragos,2019_schaerer,2019_senchyna}. The ionizing emission from post-AGB stars appears after about 100~Myr at a rate that is about five orders of magnitudes lower than that of the most massive stars \citep{2019_byler}. The accreting compact objects provide ionizing emission at a much lower rate than living stars, but their emission is also significantly harder \citeg{2013_lepo}. Since massive stars, stripped stars and blue stragglers, are expected to emit ionizing photons at the highest rates compared to these other stellar ionizing sources, we choose to start with including only these stars.

We also do not account for the stellar rotation of massive stars in our stellar population synthesis models \citep{huang_2010, ramirez_agudelo_2013} or stars spun up from binary interaction \citep{2011_eldridge,2013_de_Mink, dufton_2011}. Stellar rotation has been predicted to lead to, for example, more luminous stars, more WR stars, and stars evolving chemically homogeneously \citeg{1987_maeder,Ekstrom_2012,2007_cantiello}. Although many effects of rotation are interesting since they predict an increased emission rate of ionizing photons \citeg{Topping_2015,2012_levesque,2019_kubatova}, there is only circumstantial observational evidence for the existence of these processes (see however, \citealt{Abdul_Masih_2019}; and circumstantial evidence from e.g., \citealt{Martins_2008,Hainich_2015}; see also \citealt{Shenar_2019,Schootemeijer_2018}).

\subsubsection{Comparison to {\sc bpass}}
\label{sec:method:binary:bpass}
Both M16 and R18 use models from {\sc bpass} version 2.0, which have very similar initial conditions to our models. However, Figure \ref{fig:single_sb} shows that even with the additional emission from stripped stars and massive blue stragglers combined, the model from {\sc bpass} produces more LyC photons over the course of a Gyr than the models used in this paper. It does not appear as dramatically because of the logarithmic scale of Figure \ref{fig:single_sb}, but the most significant difference between {\sc bpass} and our model occurs between 3 and 20~Myr. This boost in ionizing emission occurs because {\sc bpass} assumes that high-mass accretor stars evolve chemically homogeneously at $Z\leq 0.004$ \citep{2011_eldridge}.

{\sc bpass} also predicts a higher, almost flat emission rate at very late times ($>$~200~Myr) \citesee{Stanway_2018}. This difference could be related to the inclusion of post-AGB stars in {\sc bpass}, and possibly our model's inclusion of gravitational settling in the atmospheres of the low-mass stripped stars that form in older stellar populations \citesee{2018_gotberg}.  Interestingly, in the most recent model from {\sc bpass} \citep[version 2.2.1]{2017_eldridge}, fewer delayed LyC photons are produced through binary interactions.

In addition, the models from {\sc bpass} version 2.0 only go down to a metallicity of Z=\num{1e-3}. This metallicity is still high for very metal poor high-redshift dwarf galaxies, and the metallicity of a model can have a significant affect on the amount of ionizing photons. For example, in our models the total fraction of ionizing photons that come from binary products over 1~Gyr increases from our lowest metallicity models to our next lowest metallicity models from 6\% to 9\%. The models we use here go down to either Z=\num{4e-4} or Z=\num{2e-4}.

\subsection{Calculating the Escape Fraction}
\label{sec:method:ray}
The rate of photons that escape the virial sphere of each halo for each star particle in that halo, when accounting for absorption by hydrogen, helium, and dust, as a function of frequency, $\nu$, is
\begin{equation}
\begin{multlined}
    \label{eq:absorb}
    \dot{N}_{\rm{esc,i}}(\nu) = \int_{0}^{4\pi} d\Omega \dot{N_{\rm{i}}}(\nu)
     \exp[-\sigma_{\rm{HI}}(\nu)N_{\rm{HI}}(\Omega) \\
    - \sigma_{\rm{HeI}}(\nu)N_{\rm{HeI}}(\Omega) 
    - \sigma_{\rm{HeII}}(\nu)N_{\rm{HeII}}(\Omega) \\
    - k_{\rm{ext}}(\nu)\Sigma(\Omega)],
\end{multlined}
\end{equation}
where $\Omega$ is the solid angle; $N_{\rm{HI}}(\Omega)$, $N_{\rm{HeI}}(\Omega)$, and $N_{\rm{HeII}}(\Omega)$, are the neutral hydrogen, and neutral and singly ionized helium column densities output from the {\sc ramses} snapshots; $k_{\rm{ext}}$ is dust extinction opacity; and $\Sigma(\Omega)$ is the surface density of dust along the line of sight also output from the {\sc ramses} simulations. $\dot{N_{\rm{i}}}(\nu)$ is the age- and metallicity-dependent number of photons per second emitted by an individual star particle, i, in a {\sc ramses} snapshot at frequency, $\nu$. The value of $\dot{N_{\rm{i}}}(\nu)$ is extrapolated from the age- and metallicity-dependent {\sc {\sc starburst99}} spectrum, with the age- and metallicity-dependent photon rates from the stripped star or massive blue straggler spectrum added on for the SSB and MBS runs, respectively.

The neutral hydrogen, $\sigma_{\rm{HI}}(\nu)$, neutral helium, $\sigma_{\rm{HeI}}(\nu)$, and singly ionized helium, $\sigma_{\rm{HeII}}(\nu)$, cross-sections are calculated as in \cite{1989_osterbrock}. The frequency-dependent dust extinction opacity, $k_{\rm{ext}}(\nu)$, is extrapolated from the dust model for the metal poor Small Magellenic Cloud in \citet{2001_weingartner_draine,2001_li_draine}.

The escape fraction for each star particle, $f_{\rm{esc,i}}$, in each halo snapshot is then computed as the ratio
\begin{equation}
    \label{eq:fesc}
    f_{\rm{esc,i}}=\frac{\int_{13.6\rm{eV}/h}^{\infty}\dot{N}_{\rm{esc,i}}(\nu)d\nu}{\int_{13.6\rm{eV}/h}^{\infty}\dot{N}_{\rm{i}}(\nu)d\nu}.
\end{equation}

For each {\sc ramses} snapshot of each halo in each run,
\begin{equation}
    \label{eq:fesc_mean}
    f_{\rm{esc}}=\frac{\sum_if_{\rm{esc,i}}\dot{N}_{\rm{i}} }{\sum_i\dot{N}_{\rm{i}}},
\end{equation}
where $f_{\rm{esc}}$ is the photon production rate-weighted mean escape fraction for each halo snapshot.

Note that our calculation of the number of escaping ionizing photons is done in post-processing using a simulation that did not include the effects of binary evolution. As a result, we predict our value of $f_{\rm{esc}}$ is likely lower than it would be if it were calculated directly in the cosmological simulation. We expect that $f_{\rm{esc}}$ is lower because when we include photons from binary stars only in post-processing, photons from this population that fail to escape also do not ionize the gas that absorbs them. If these photons were included in the cosmological simulation they would ionize the gas that absorbs them, helping to clear a path for future photons. It is also useful to note that because our calculation is done in post-processing the total number of escaping LyC photons from both additional sources, stripped stars and massive blue stragglers, combined is simply the sum of the number of escaping LyC photons from each source calculated separately.

\begin{figure*}
\includegraphics[width=\textwidth]{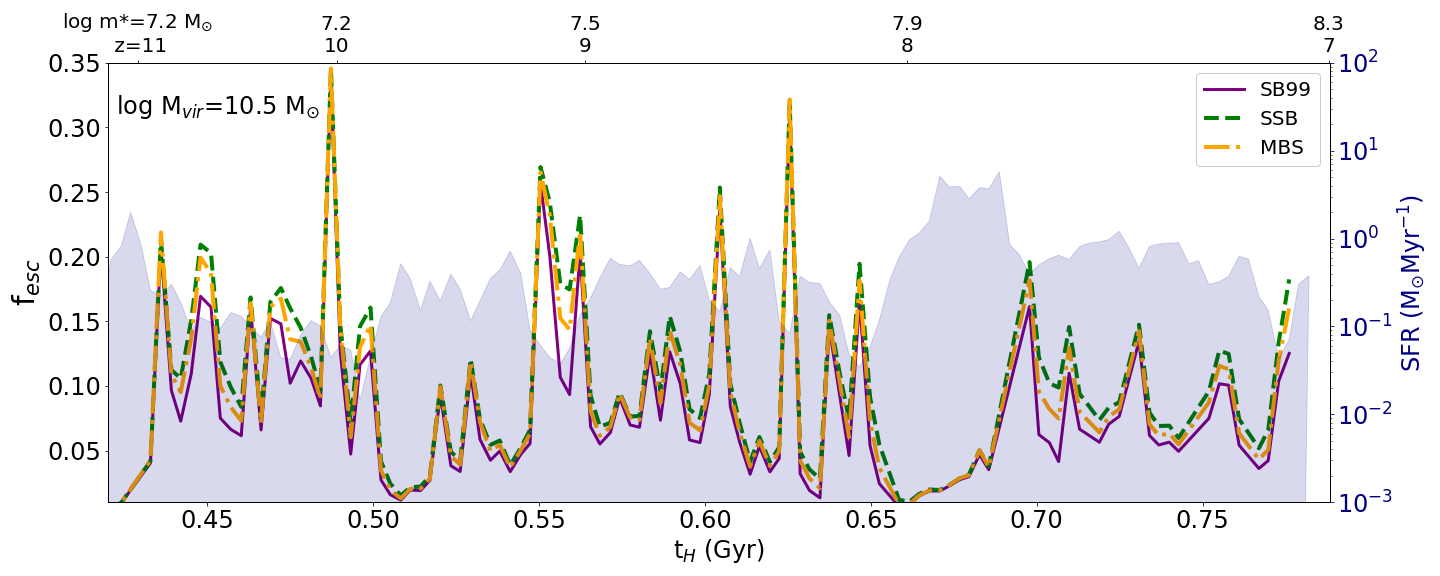}
\caption{Evolution of the escape fraction ($f_{\rm{esc}}$) for an example halo as a function of the age of the Universe in Gyr. The SB99 run is shown in purple (solid line), the SSB run in green (dashed line), and the MBS run in orange (dashed-dotted line). The purple shaded region shows the star formation rate (SFR) in $\rm{M_{\odot}~yr^{-1}}$. Redshifts corresponding to the ages of the Universe shown on the lower x-axis are given on the upper x-axis along with the log of the stellar mass (in $\rm{M_{\odot}}$) of the halo at that redshift. As noted in KC14, the peaks in $f_{\rm{esc}}$ are out of phase with the SFR.}
\label{fig:kimm14}
\end{figure*}

\begin{figure*}
\includegraphics[width=\textwidth]{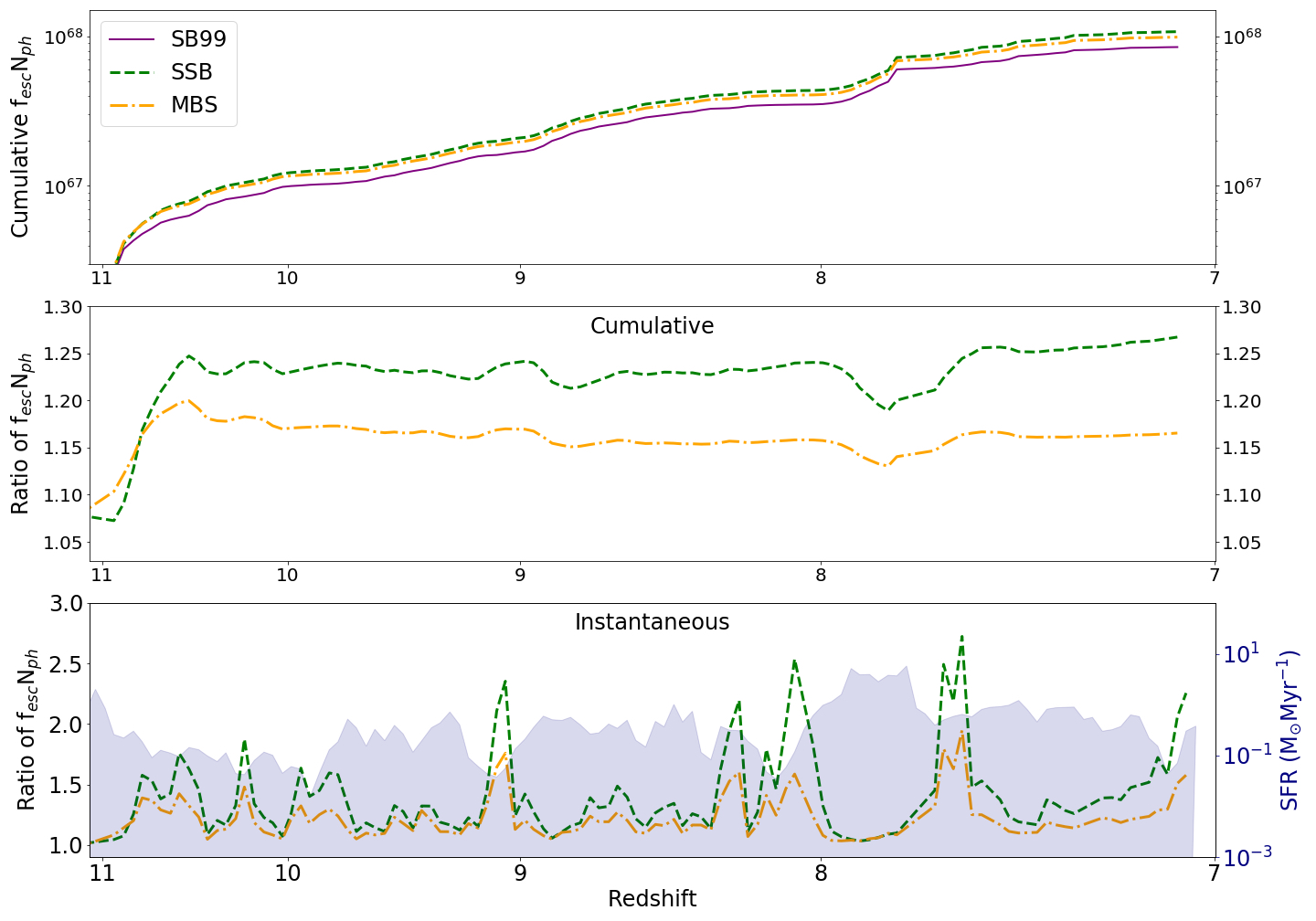}
\caption{Comparison of the number of escaping LyC photons as a function of redshift for the different runs for 
the halo shown in Figure \ref{fig:kimm14}. 
The top panel shows the cumulative amount of LyC photons escaping the virial radius of this halo using the same color-scheme and line-style as in Figure \ref{fig:kimm14}. The middle panel shows 
the ratio of cumulative, escaped LyC photons for the SSB run to the SB99 run, in green, and that of escaped LyC photons for the MBS run to the SB99, in orange. 
The bottom panel is analogous to the middle panel but
for the instantaneous escape fraction. We have also plotted the star formation rate (SFR) as in Figure \ref{fig:kimm14} to show the offset between when the increase in the number of escaping photons is greatest, and when the SFR is highest.
Looking forward, we note that these ratios are larger than the mean values shown in Figure \ref{fig:nph_fesc}.}
\label{fig:oneH_big}
\end{figure*}

\begin{figure*}
\includegraphics[width=\textwidth]{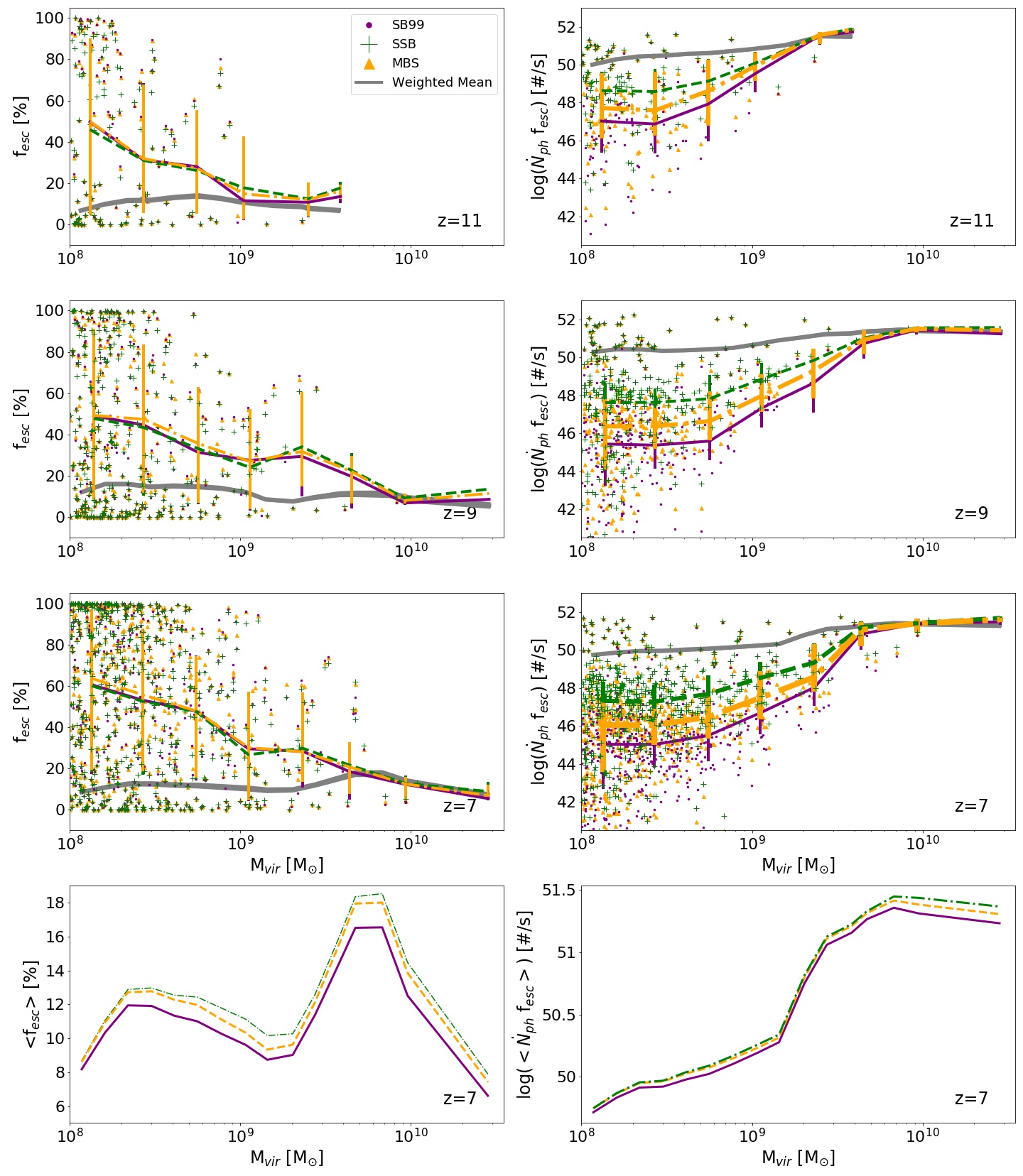}
\caption{The escape fraction ($f_{\rm{esc}}$) in percent (left panel) and the logarithm of the photon production rate multiplied by the escape fraction in s$^{-1}$ (right panel) as a function of virial mass in M$_{\rm{\odot}}$ for all halos at a redshift of 11 (top panel), 9 (middle panel), and 7 (bottom panel). The color-scheme and line-styles are the same as in Figure \ref{fig:kimm14}. In the top three panels each point represents one halo at that redshift, and the colored lines represent the unweighted (by photon production rate) median values, with the error bars showing the interquartile ranges. The grey line shows the values for the box-car smoothed photon-weighted means. The thickness of the grey line corresponds to the difference between the SB99 run and the SSB run, i.e. the top of the line is the value of the photon-weighted mean for the SSB run and the bottom of the line is the value of the photon-weighted mean for the SB99 run. The bottom panel zooms in on the values of the photon-weighted mean escape fraction and production rate of escaping photons for the three different runs at $z=7$. To increase statistical significance we combine the results over six consecutive snapshots at each redshift.}
\label{fig:scatter}
\end{figure*}

\begin{figure*}
\includegraphics[width=\textwidth]{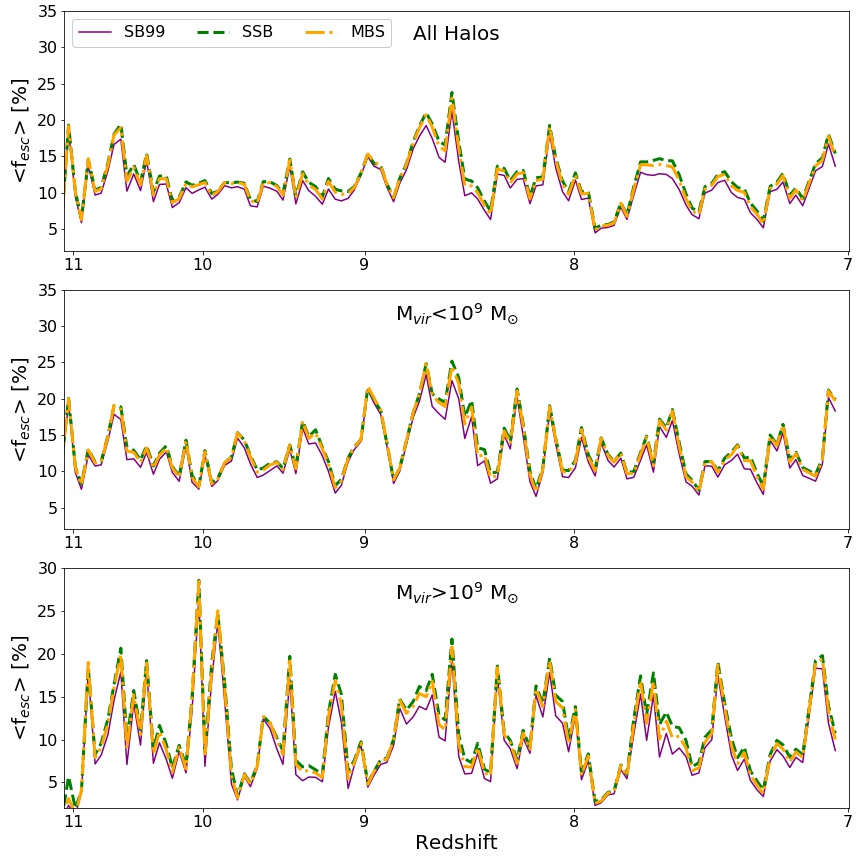}
\caption{The photon production rate-weighted mean escape fraction ($\langle f_{\rm{esc}} \rangle$), in percent, as a function of redshift. The top panel shows $\langle f_{\rm{esc}} \rangle$ for all halos in the simulation. The middle panel shows $\langle f_{\rm{esc}} \rangle$ for halos with virial masses less than $10^9$~M$_{\rm{\odot}}$ and the lower panel shows $\langle f_{\rm{esc}} \rangle$ for halos with virial masses greater than $10^9$~M$_{\rm{\odot}}$. The color-scheme and line-styles are the same as in Figure \ref{fig:kimm14}. $\langle f_{\rm{esc}} \rangle$ is greater for both runs that include binary effects than for the SB99 run over all redshifts in every panel.}
\label{fig:mass_group}
\end{figure*}

\begin{figure}

\includegraphics[width=0.5\textwidth]{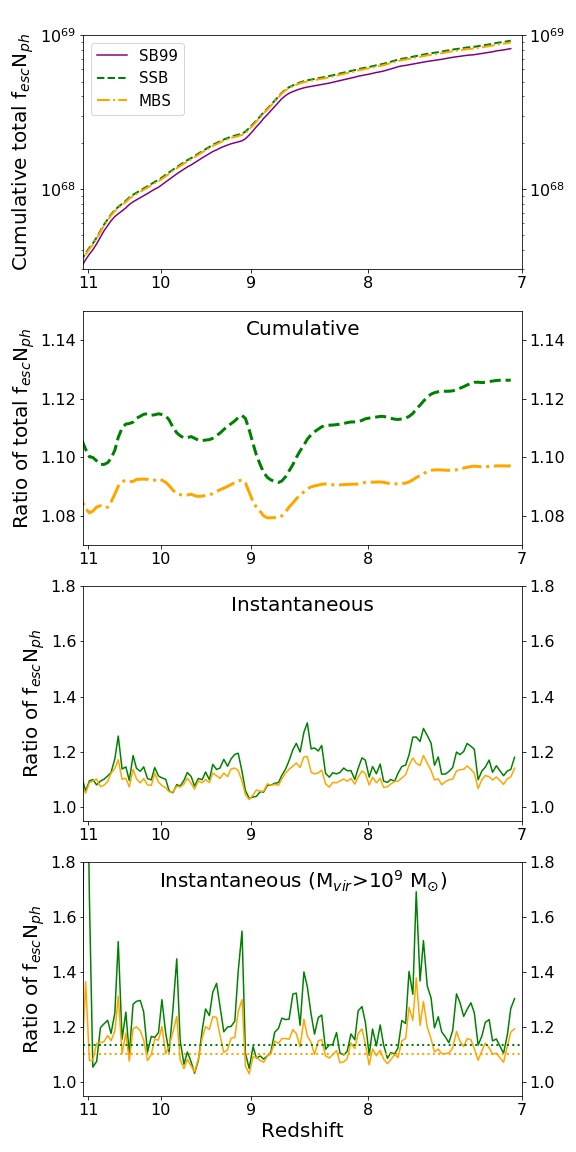}
\caption{The top panel shows the cumulative total number of LyC photons that escape the virial radius of their halo for all halos as a function of redshift. The second panel from the top shows the ratio between the total cumulative number of escaping LyC photons for the SSB/MBS runs and the SB99 run as a function of redshift. The third panel from the top and the bottom panel show the instantaneous ratio between the number of escaping LyC photons for the SSB/MBS runs and the SB99 run as a function of redshift for all halos and halos above $10^9$~M$_{\rm{\sun}}$, respectively. The color-scheme and line-types are the same as in Figure \ref{fig:kimm14}. The dotted lines in the bottom panel show the average instantaneous ratio for all of the halos.}
\label{fig:nph_fesc}
\end{figure}

\section{Results}
\label{sec:results}

\subsection{An Example Halo}
\label{sec:results:one_halo}

Before presenting statistical results,
we use one halo to illustrate some 
basic effects on the escape fraction of ionizing photon 
due to binary evolution in
Figures \ref{fig:kimm14} and \ref{fig:oneH_big}.
A relatively massive halo,
for which we are able to construct a merger
tree to a very high redshift, is used here as an example.
The virial and stellar mass of this halo at $z=7$ is \num{3.16e10}~M$_{\rm{\odot}}$ and \num{2e8}~M$_{\rm{\odot}}$, respectively.

KC14 found that more massive halos, on average,
have lower escape fractions. The photon production rate-weighted escape fraction of this halo for the SB99 run, is 11\%. 
When photons from either stripped stars or massive blue stragglers are included,
we see that the escape fraction increases significantly.

Figure \ref{fig:kimm14} shows the escape fraction as a function of time, in Gyr on the bottom axis and redshift on the top axis. The SB99 run is shown in purple (solid line), the SSB run in green (dashed line), and the MBS run in orange (dashed-dotted line). This color scheme (and line type) is consistent for all figures in this section. The shaded purple region shows the star formation rate (SFR) in M$_{\rm{\odot}}$~yr$^{-1}$ as a function of time, which is calculated by summing over star particles formed in the last 1 Myr. The logarithm of the stellar mass of the halo at each redshift is also shown on the upper axis above the redshift.

As shown first in KC14, the escape fraction decreases significantly when the SFR is highest due to the presence of large amounts of gas during star formation which occurs deep in the cores of giant molecular clouds. There is therefore a delay between peak star-formation and the increase in the escape fraction that occurs once much of the birth cloud has been cleared through supernova-driven blastwaves. This delay is commonly seen
in dwarf galaxy simulations at high redshift \citeg{2009_wise_cen,2014_wise,2015_ma}. For this halo the escape fraction is always higher for both runs that include products of binary interactions than for the run that does not. 

The largest increases in the escape fraction in the SSB and MBS runs compared to the SB99 run occur about 10 Myr after a major starburst has ended, for example at $t_H\sim 0.705$~Gyr in Figure \ref{fig:kimm14}. At first the escape fractions of all three runs increase, because supernova feedback 
has significantly cleared out the birth clouds
of the starburst. However, shortly after the escape fraction for the SB99 run increases, it will rapidly decrease, because after 10~Myr the most massive single-stars are gone. The escape fractions of the SSB and MBS runs decrease less, because LyC photon production for stripped stars and massive blue stragglers peaks at around 10~Myr.

Figure \ref{fig:oneH_big} compares the number of escaped photons as a function of redshift for the three runs. The top panel shows the cumulative number of escaped photons for SB99 (solid purple),
SSB (dashed green) and MBS (dot-dashed orange).
The middle panel shows the ratio 
of the cumulative number of photons of SSB+SB99 to SB99 (dashed green),
and that of MBS+SB99 to SB99 (dot-dashed orange),
respectively.
The bottom panel is similar to the middle panel 
but for the instantaneous number of photons. The shaded purple region once again shows the SFR in M$_{\rm{\odot}}$~yr$^{-1}$.
We see that 
the instantaneous ratio of escaped photons changes significantly over time, often reverting back to close to order unity directly after a starburst. 
For this example halo,
we see that when stripped stars are included, 27\% more LyC photons escape into the IGM by $z=7$, compared to when only massive stars are accounted for. When massive blue stragglers are included 17\% more LyC photons escape. Combined LyC emission increases by 42\% for this halo when both stripped stars and massive blue stragglers are included. This would increase the escape fraction of this halo to $\sim$14\% when stripped stars are included, and $\sim$16\% when both stripped stars and massive blue stragglers are included. The dip present at $\rm{z}\sim 7.7$ is from the boost in the {\sc starburst99} ionizing emission rate right after a new starburst.

\subsection{Statistical Results}
\label{sec:results:all}

The cosmological simulations of KC14 include hundreds of halos ranging in virial mass from $10^8$~M$_{\rm{\odot}}$ to $10^{11}$~M$_{\rm{\odot}}$ at $z\sim7$. The left panels of Figure \ref{fig:scatter} show the escape fraction ($f_{\rm{esc}}$), in percent, as a function of virial mass. The right panels show the production rate of ionizing photons that escape the virial sphere (in number per second) as a function of virial mass. In the top three panels each point represents an individual halo. The purple, green, and yellow lines show the median values for all halos within a mass range as a function of the median of those mass ranges for the SB99, SSB, and MBS runs, respectively. The error bars represent the interquartile range. The grey lines show the box-car smoothed photon-weighted mean values for the three runs. The bottom of the grey line represents the mean for the SB99 run and the top of the line represents the mean of the SSB run. The bottom panel is a zoom-in to show the box-car smoothed photon-weighted means for the three runs at $z=7$ with the same color-scheme as in Figure \ref{fig:kimm14}. To increase the statistical significance, results from six consecutive snapshots are combined to determine the medians, interquartile ranges, and means for each redshift specified. The color scheme and line-types are the same as above. The top panels are at $z=11$, the panels directly below them are at $z=9$, and the bottom two panels are at $z=7$.

We find a median unweighted escape fraction of roughly 10\%-50\% for all halo masses, although there is a large interquartile range. Less massive halos ($M_{\rm{vir}}<10^9~M_{\rm{\odot}}$) overall have a higher median escape fraction than higher mass halos, as noted earlier in KC14. There is not a large variance in the escape fraction over different redshifts, except an increased smoothness in the median values as there are more halos at later redshifts.  

The median production rate of escaping photons increases with virial mass, as more massive galaxies tend to have higher SFRs. The median production rates also decrease slightly with decreasing redshift, because for a given halo the SFR decreases and the metallicity increases with redshift. The interquartile range is large for the production rate of escaping photons too, but the median values range from roughly $10^{45}$~s$^{-1}$ at lower virial masses to $10^{52}$~s$^{-1}$ at higher virial masses.

In Figure \ref{fig:scatter}, the median unweighted escape fraction is slightly lower for the SSB run than the SB99 run for halos with virial masses less than about $10^9~$M$_{\rm{\odot}}$, but higher for the most massive halos. The median unweighted escape fraction for our MBS run is very similar to the median unweighted escape fraction for the SB99 run for halos with virial masses less than about $10^9~$M$_{\rm{\odot}}$, and then slightly higher for the more massive halos. The rate of escaping photons is greater at all redshifts for all mass halos for the SSB and MBS runs, but by a larger factor for less massive halos than more massive halos. At low masses this factor is around 40-200 for the SSB run and around 5-10 for the MBS run, and at high mass this factor is around order unity to 20 for the SSB run and 1-5 for the MBS run. These factors are slightly greater at later redshifts than earlier ones. 

The differences in the median rate of escaping photons between the SSB and MBS runs and the SB99 run are greater at lower masses because the starbursts of low mass halos are more episodic, meaning that many halos at these low masses have undergone very little recent star formation. During periods of low star formation, stripped stars, and to a lesser extent massive blue stragglers, become the dominant source of LyC radiation, because of the relatively long time range over which these sources have a significantly higher LyC photon rate. For example, in Figure \ref{fig:single_sb} the {\sc starburst99} photon rate of a $10^6~$M$_{\rm{\odot}}$ starburst with an age of 100~Myr will be only $\sim10^{47}$~s$^{-1}$. Meanwhile, at that age stripped stars will have a photon production rate roughly 200 times greater, which accounts for the two orders of magnitude difference between the escaping photon rates of the SSB and SB99 runs in low mass halos.

Stripped stars in particular have harder ionizing emission than massive stars, and so these low mass galaxies whose LyC emission is dominated by stripped stars would also emit harder spectra. These low mass galaxies emitting harder spectra may be observable in the future by the James Webb Space Telescope (JWST), or even produce a unique signature in future observations of reionization bubbles by the Nancy Grace Roman Space Telescope. Interestingly, if JWST observes these low mass halos then the observed ionizing emission would almost exclusively come from binary products, dominated by stripped stars (see right panel of Figure \ref{fig:scatter}), which means that these low mass halos could be very useful for studying binary interactions at high redshift. However, the dramatic increase in the photon rate for the SSB run for the bursty low mass halos is also responsible for the slight decrease in the median escape fractions for the SSB run. Star particles which were not emitting any hydrogen-ionizing photons previously, now emit some radiation that can still be trapped, which lowers the escape fraction.

The box-car smoothed photon-weighted mean escape fractions, shown as the grey line in the top three left-hand panels of Figure \ref{fig:scatter}, are mostly lower than the median escape fractions. The most massive halos occasionally have similar mean and median escape fractions because there are not many of them. The difference between the unweighted median and weighted mean escape fractions is not surprising since, as KC14 found and as can be seen in Figure \ref{fig:kimm14}, star formation peaks are out of phase with peaks in the escape fraction by 10-200~Myr. The effects of the star formation peak and escape fraction peak phase difference are somewhat mitigated by the delayed photons in the SSB and MBS runs leading to the higher weighted mean escape fractions for these runs. These higher escape fractions can be seen in the bottom panel of Figure \ref{fig:scatter}. Photons from the stripped star component are particularly delayed, with the photon rate for this component remaining above $10^{49}$~s$^{-1}$ for over $\sim 100$~Myr after the initial starburst.

The bottom right panel of Figure \ref{fig:scatter} shows that at $z=7$ the photon-weighted mean production rate of escaping photons is greater for both runs that include binary effects, although the dramatic increase in the production rate for lower mass halos is gone. In fact, the difference in $\langle \dot{N}_{\rm{ph}}f_{\rm{esc}}\rangle$ is largest for the most massive halos. $\langle \dot{N}_{\rm{ph}}f_{\rm{esc}}\rangle$ for the lower mass halos does not increase as dramatically between runs because these halos' photon rates are so much higher during the 20~Myr after a starburst that photons from young stellar populations completely dominate the mean rate of escaping LyC photons. Nonetheless, the fact that most low mass galaxies will have their LyC emission rate so enhanced during a majority of their evolution while their SFRs are low provides an exciting opportunity to learn about binary products if these galaxies could be observed.

Figure \ref{fig:mass_group} shows the photon production rate-weighted mean escape fractions as a function of redshift, for all halos in our simulation (upper panel), halos with a virial mass less than $10^9$~M$_{\rm{\odot}}$ at that redshift (middle panel), and halos with a virial mass greater than $10^9$~M$_{\rm{\odot}}$ at that redshift (lower panel). The color-scheme and line-types are the same as in Figure \ref{fig:kimm14}. Overall, the mean escape fractions stay mainly within 10 and 20 percent, with higher mass halos 
having a somewhat larger variance.
The large temporal variations are dominated by a handful of large starburst events. The weighted mean escape fraction is greater at all redshifts for both the SSB and the MBS runs than the SB99 run.

Figure \ref{fig:nph_fesc} compares the total number of photons which escape the virial sphere for all halos as a function of redshift for the various runs. The top panel shows the cumulative number of escaping LyC photons for all halos in each run. The panel directly below it shows the ratio of these numbers of escaping photons for the SSB and MBS runs to the SB99 run in green and orange, respectively. The bottom two panels show the instantaneous ratios of the total number of escaping photons for all halos (third panel) and halos more massive than $10^9$~M$_{\rm{\sun}}$ (bottom panel). The color-scheme and line-types are the same as in Figure \ref{fig:kimm14}. The dotted line in the bottom panel shows the average instantaneous ratio for all the halos, for comparison.

The inclusion of stripped stars and massive blue stragglers increases the total number of escaped photons at a redshift of $\sim7$ by around $10^{68}$ photons and \num{8e67} photons, respectively, that is by a factor of $\sim1.125$ and $\sim1.10$, respectively. Including both binary products would increase the total number of escaped photons by $\sim$~\num{2e68}, or by 22.5\%. After the initial high redshift starbursts, the overall trend is for the ratio of escaped photons between the SSB and MBS runs and the SB99 run to increase as redshift decreases. This trend is also present in the ratios of the median production rate of escaping photons, as mentioned above in the discussion of Figure \ref{fig:scatter}. This ratio increases because the SFR decreases with redshift. A lower SFR means there will be more older stellar populations at lower redshifts, and stripped stars and massive blue stragglers are the dominant sources of LyC radiation in these populations. If the escape fractions for the various runs were identical, then for star particles at a metallicity of \num{2e-4} the additional photons from stripped stars and massive blue stragglers would lead to an increase in the total number of photons by a factor of 1.036 and 1.033, respectively. Therefore, LyC photons from stripped stars and massive blue stragglers are significantly more effective at escaping their host galaxies. In \S \ref{sec:compare} using Equation \ref{eq:fesc_us} we calculate that the mean escape fraction for photons from these binary products is 57\%, versus 14\% for the massive stars.

The instantaneous ratios between the SSB/MBS and SB99 runs of the total number of escaping ionizing photons at a given redshift vary between just above unity to 1.3 for all halos. When only considering halos more massive than $10^9$~M$_{\rm{\sun}}$, these same ratios vary between just below 1.1 and 1.8, from $z=11$ to $z=7$. The instantaneous ratios of the more massive halos often staying above the mean value of the instantaneous ratio for all the halos (shown in the bottom panel as the dotted line). In the bottom, right panel of Figure \ref{fig:scatter} $\langle \dot{N}_{\rm{ph}}f_{\rm{esc}}\rangle$ also increases the most for runs including binary products compared to the SB99 run for higher mass halos. One reason these more massive halos will have a larger increase in the number of escaping photons is because larger halos which undergo more starbursts will have more stellar populations at ages of around 10-30~Myr since the starburst. Stellar populations at these ages will still have high enough photon rates to influence the total number of escaped photons, as well as significant and even dominant contributions from stripped stars and massive blue stragglers.

Additionally, KC14 found when a single-stellar population synthesis model is used the most massive halos tend to have lower escape fractions. They attribute this to the fact that in larger halos young massive stars can be buried in many star-forming clouds that are more resilient to SN feedback arising in neighboring star clusters. Our results suggest that delayed photons have a greater chance of escaping these halos because they allow more time for additional SN feedback to occur, not just in the cloud where the radiation is coming from, but also in adjacent clouds which will have more time to undergo their own stellar evolution that will generate feedback.

The massive example galaxy shown in Figures \ref{fig:kimm14} and \ref{fig:oneH_big} (see 
\S \ref{sec:results:one_halo}) clearly has a greater than average increase in escaped LyC photons when stripped stars and massive blue stragglers are included. Because it is one of the more massive galaxies in the simulation it fits the overall trend that more massive galaxies have a larger additional amount of escaping photons in our SSB and MBS runs.

\section{Comparison to Previous Studies}
\label{sec:compare}

M16 used the stellar population synthesis models from {\sc bpass} that include binary evolution to calculate the escape fraction of LyC for three example mock halos. For a high mass ($M_{\rm{vir}}=10^{10}~M_{\rm{\odot}}$) halo, similar in mass to our example halo in 
\S \ref{sec:results:one_halo}, M16 found that including binary evolution increased their escape fraction by a factor of $\sim 3$ from $\sim5$\% to $\sim14\%$. For our SB99 run, the more massive halos, like our example halo, had smaller than average escape fractions of $\sim 11\%$. Combining the increase in the number of escaped photons from both the SSB and MBS runs for our example halo increases the total effective escape fraction by a factor of 1.44 from 11\% to 16\%. 

R18 also used models from {\sc bpass}. They found that adding binary stellar evolution to their simulations increased the luminosity-weighted mean escape fraction for their large sample of halos by a similar factor to M16 of $\sim 3$ from $f_{\rm{esc}}=2.7\%$ to $f_{\rm{esc}}=8.5\%$. The photon-rate weighted mean escape fraction over all of our simulated halos increased by a factor 1.225 from 14\% to 17\%.

Our single-stellar population synthesis run clearly produces significantly larger values of $f_{\rm{esc}}$ than the single-stellar population synthesis runs in M16 and R18. One reason for this is that our single-stellar population model includes runaway O/B stars (see 
\S \ref{sec:method:runaways}), which KC14 showed increased their overall value of $f_{\rm{esc}}$ from 11\% to 14\%. However, 11\% is still a factor of 2-5 larger than the percent of photons escaping in the M16 and R18 simulations. This discrepancy is likely due to the different star formation and feedback schemes among the different simulations. 

R18, in particular, use a varying star formation efficiency (star formation efficiency per dynamic time) that leads to preferential star formation in higher density regions. Because our simulations use a fixed star formation efficiency of 2\% per dynamical time, even though the density threshold for star formation in their simulation is lower than ours (10~cm$^{-3}$ in theirs vs 100~cm$^{-3}$ in ours), stars in our simulations are less abundant in the densest regions. These denser regions of gas where the majority of stars are forming in the R18 {\sc sphinx} simulations are more difficult to clear with stellar feedback, which is probably the reason R18 has much lower values for their escape fractions with and without binary stellar populations, exacerbated by the omission of runaway
O/B stars in their simulations.

Importantly, the galaxy stellar mass and luminosity functions in R18's simulations 
are consistent with observations but 
lie near the upper end of observational constraints and other recent models.
Their results would still fall within the
observational constraints even 
if the amplitudes of the stellar masses were decreased by as much as a factor of 
two through increased SN feedback. This stronger SN feedback would lead to an increase in $f_{\rm{esc}}$ and a decrease in star formation, and therefore ionizing luminosities. These changes would make the results of R18 more similar to ours here.

The different star formation treatment in R18 is also in part responsible for their much larger increase in $f_{\rm{esc}}$ when including binary stellar evolution in their simulation. Because the birth clouds in R18 are denser, it takes longer for feedback and LyC photons to clear them. Therefore, most massive stars will be gone by the time these clouds are cleared, and so very few LyC photons from this stellar population are able to escape. Delayed LyC photons from binary stellar evolution, on the other hand, will be able to escape. As a result the increase in the escape fraction due to binary stellar evolution will be greater. The denser birth clouds in R18 also
allow for a concerted launch of SN-driven blastwaves from a more concentrated, larger
stellar cluster with nearly coeval
star formation, leading to a ``cleaner" sweep
of the remaining gas in star birth clouds.
This can be seen indirectly in the rather large
changes in escape fraction for singles and binaries
in Figure 13 of R18.

Most significantly, the escape fractions in M16 and R18 both increase with the inclusion of binary evolution by a greater factor than in this paper because the number of additional LyC photons from binary stellar evolution is greater in both of those papers than in this paper. The number of LyC photons increases by roughly 50\% and 70\% for M16 and R18, respectively, while here we only have a 6\% increase. Figure \ref{fig:single_sb} shows that the LyC photon ionization rate of the lowest metallicity model from {\sc bpass} that M16 and R18 use for their runs that include binary products is significantly greater than ours. As discussed in \S \ref{sec:method:binary:bpass}, this difference is likely mainly due to the chemically homogeneously evolving accretor stars in {\sc bpass}. The high LyC photon rate in R18, may be the reason why they find that their simulations are reionized slightly earlier than observations would predict, despite a low escape fraction of 7\%.

Because the M16 and R18 simulations have a greater increase in the number of LyC photons due to binary interaction, the escape fractions of the additional ionizing photons from these interacting binaries, $f_{\rm{esc,bin}}$, are more heavily weighted in their overall escape fractions than in ours. For example, 40\% of the ionizing radiation in the R18 simulation that uses the spectrum from {\sc bpass} comes from interacting binaries. Therefore, if $f_{\rm{esc,ss}}$ is the escape fraction for the single star component, the overall escape fraction for the R18 simulation that uses the spectrum from {\sc bpass} is,
\begin{equation}
\label{eq:fesc_R18}
    [f_{\rm{esc}}]_{\rm{R18}} = 0.6f_{\rm{esc,ss}} + 0.4f_{\rm{esc,bin}}.
\end{equation}
On the other hand, in this paper only 7\% of the ionizing radiation comes from interacting binaries. The overall escape fraction here is,
\begin{equation}
\label{eq:fesc_us}
    f_{\rm{esc}}= 0.93f_{\rm{esc,ss}} + 0.07f_{\rm{esc,bin}}.
\end{equation}
Clearly $f_{\rm{esc,bin}}$ is more heavily weighted in R18.

It is therefore useful to compare $f_{\rm{esc,bin}}$, among the three papers. Physically, $f_{\rm{esc,bin}}$ represents the escape fraction of the additional LyC radiation from interacting binaries. We can calculate $f_{\rm{esc,bin}}$ using a rearranged generalized version of equations \ref{eq:fesc_R18} and \ref{eq:fesc_us},
\begin{equation}
    \label{eq:fesc_gen}
    f_{\rm{esc,bin}}= \frac{f_{\rm{esc}}-xf_{\rm{esc,ss}}}{1-x},
\end{equation}
where $x$ is the fraction of LyC radiation from the single-stellar component in simulations that include binary interaction. From each paper we know $x$; the escape fraction of the run which includes only the single-stellar population, which we take as $f_{\rm{esc,ss}}$; and the overall escape fraction of the run that includes binary evolution, $f_{\rm{esc}}$. This information will give accurate values of $f_{\rm{esc,bin}}$ for this paper and M16 because both apply a ray tracer in post-processing. However, using Equation \ref{eq:fesc_gen} to calculate $f_{\rm{esc,bin}}$ for R18 is not entirely accurate because they use a model from {\sc bpass} in the cosmological simulation itself. Implementing a model from {\sc bpass} in the cosmological simulation itself means that the extra photons from binary products can also help LyC photons from single stars to escape.

With this caveat for the value of $f_{\rm{esc,bin}}$ for R18 in mind, $f_{\rm{esc,bin}}=32\%,17\%,57\% $ for M16, R18, and this paper, respectively. Both values of $f_{\rm{esc,bin}}$ for M16 and R18 are greater than their values of $f_{\rm{esc,ss}}$ by a factor of $\sim$ 6, which one would expect given that the radiation from single stars is so well trapped in M16 and R18. Our value increases from $f_{\rm{esc,ss}}$ to $f_{\rm{esc,bin}}$ by a slightly smaller factor of $\sim$4. $f_{\rm{esc,bin}}$ in this paper is much greater than the values from both M16 and R18 once again because of the different feedback and star formation schemes of the various simulations. Interestingly, \cite{Gotberg_2020} found that if the escape fraction of stripped stars is high, $\sim 50-80\%$, they can provide most of the LyC emission required for reionization, despite massive stars having typical escape fractions of $\sim 10-20\%$. Our simulations support that the high escape fraction \citet{Gotberg_2020} set in their simulations is achievable for binary products.

\section{Conclusions}
\label{sec:discuss}

\begin{figure}
\includegraphics[width=0.5\textwidth]{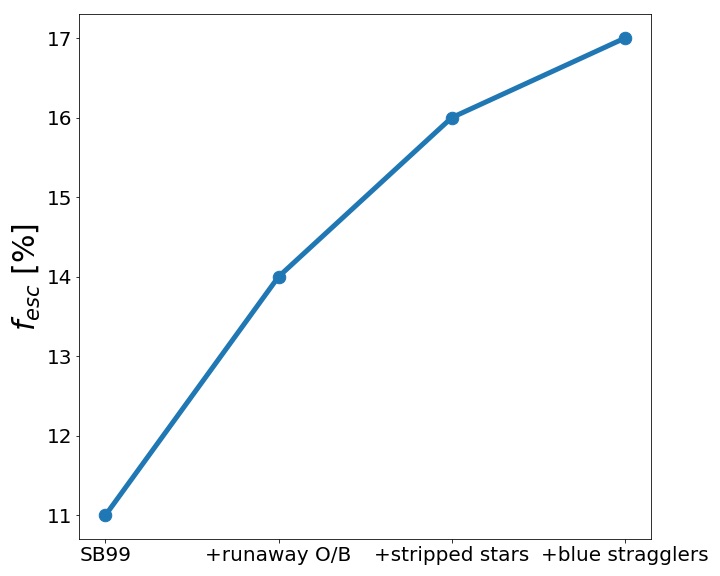}
\caption{Photon production rate-weighted mean escape fractions calculated from the {\sc ramses} cosmological simulation run by KC14 when using a {\sc {\sc starburst99}} spectrum, and when adding in runaway O/B stars, then stripped stars, then massive blue stragglers. The final two steps are what we examine in this paper and they increase the effective escape fraction from 14\% to 17\%.}
\label{fig:summary}
\end{figure}

Using ultra-high resolution cosmological radiation-hydrodynamic simulations, we study the effects of interacting binaries on the fraction of LyC photons that escape from their host galaxy. Binary evolution modeling inevitably remains uncertain due to lack of direct information of stellar binarity and the stellar initial mass function in galaxies at the epoch of reionization. However, utilizing results from our simple model of blue stragglers and state-of-the-art model of stripped stars based on local stellar observations, we demonstrate that the extra LyC photons from these two sources increase $f_{\rm{esc}}$
significantly. 

LyC photons from these two sources are delayed relative to the initial starbursts, on time scales of $\sim 10$~Myr. As a result the photons released from stripped stars and massive blue stragglers increase the number of escaped photons by a more significant factor than just the number of additional photons they produce.
For example for a single $10^6$~M$_{\rm{\odot}}$ starburst at a metallicity of $\lesssim \num{2e-4}$ our SSB model produces a factor of 1.036 more photons than the SB99 model, and the MBS model produces a factor of 1.033 more photons than the SB99 model. These factors are significantly less than the factor of 1.125 and 1.10 additional escaping photons by $z=7$ we see in our SSB and MBS runs, respectively. The physical reason is that the escape fraction and the instantaneous SFR are often strongly anti-correlated with a phase shift of $10-200$~Myr, reflecting the fact star formation tends to occur in regions of high obscuration and that it takes roughly $10-200$~Myr to clear the birth giant molecular cloud.

Our results also suggest that LyC photons from massive blue stragglers and in particular stripped stars significantly dominate the ionizing emission of low mass galaxies for a majority of these galaxies' histories. If these galaxies can be detected by JWST, they would make excellent laboratories for studying interacting binaries as a function of redshift. Future work is needed to better predict the spectral shape of these galaxies, as, for example, LyC radiation from stripped stars tends to be harder than typical massive stars,
potentially masquerading as Pop III stars.

Figure \ref{fig:summary} summarizes the increase of the photon-weighted mean escape fraction due to the addition of photons from binary products from the 14\% found by KC14, who used a {\sc starburst99} spectrum and included runaway O/B stars in their simulations, to 16\% and 17\% when stripped stars and both stripped stars and massive blue stragglers are included.

The requirement for cosmological reionization remains somewhat unclear due to a combination of uncertain factors, including the initial mass function and clumping factor of the IGM as a function of redshift. R18 found that an escape fraction as low as 7\% would be enough to reionize the universe by $z\sim 6$. However, 20\% is often considered to be the minimum required escape fraction to balance the recombination rate at $z\sim 6$, if the initial mass function at $z\sim 6$ is not too different from its local counterpart and a clumping factor of $\sim 6$ is adopted \citeg{2012_kuhlen_FG,2015_robertson}.
While 17\% is still a few points lower than the quoted 20\%, the most important implication is that our modeling that includes new stellar physics, high resolution,
and advanced treatment of supernova feedback, puts the stellar reionization picture
on a more solid footing. Future work is still needed to better understand the connection between the escape fractions of high redshift dwarf galaxies with their lower redshift counterparts.

\acknowledgments
AS would like to thank Charles Emmett Maher for useful discussions. AS is supported by the National Science Foundation Graduate Research Fellowship Program under Grant No. \#DGE1656466. TK was supported in part by the National Research Foundation of Korea (NRF-2017R1A5A1070354 and NRF-2020R1C1C100707911) and in part by the Yonsei University Future-leading Research Initiative (RMS2-2019-22-0216). YG acknowledges the funding from the Alvin E.\ Nashman fellowship for Theoretical Astrophysics.
The research is supported in part by NASA grant 80NSSC18K1101.
Computing resources
were provided in part by the NASA High- End Computing
(HEC) Program through the NASA Advanced Supercomputing
(NAS) Division at Ames Research Center and in part by
Horizon-UK program through DiRAC-2 facilities.

\bibliographystyle{aasjournal}
\bibliography{fesc.bib}

\end{document}